\documentclass[10pt,letter,twoside]{rmaa-rho-class/rmac-rho}
\RMxAAtemplatetype{\RMxAC} 

\vol{60}
\pages{30-36}
\thisyear{2025}

\doi{\href{https://www.astroscu.unam.mx/RMxAC/RMxAC..XX-X}{https://www.astroscu.unam.mx/RMxAC/RMxAC..XX-X}}

\usepackage{background}
\usepackage{lipsum}
\definecolor{textcolor}{HTML}{F9C051}

\title{NOCTURNE. II. Extreme radio variability in the heart of early-stage active galactic nuclei}

\author[1]{Emilia Järvelä \orcidlink{0000-0001-9194-7168}}
\author[2]{Patrizia Romano \orcidlink{0000-0003-0258-7469}}
\author[3,4]{Anne Lähteenmäki \orcidlink{0000-0002-0393-0647}}
\author[5]{Marco Berton \orcidlink{0000-0002-1058-9109}}

\affil[1]{Department of Physics and Astronomy, Texas Tech University, Box 41051, Lubbock, 79409-1051, TX, USA}
\affil[2]{INAF - Osservatorio Astronomico di Brera, Via Emilio Bianchi 46, I-23807 Merate (LC), Italy}
\affil[3]{Aalto University Metsähovi Radio Observatory, Metsähovintie 114, FI-02540 Kylmälä, Finland}
\affil[4]{Aalto University Department of Electronics and Nanoengineering, P.O. Box 15500, FI-00076 AALTO, Finland}
\affil[5]{European Southern Observatory (ESO), Alonso de Córdova 3107, Casilla 19, Santiago 19001, Chile}


\received{\today}
\accepted{\today}

\leadauthor{Järvelä et al.}

\newcommand\Text{\small ``ZIHUAGN Redemption: central kpc region'', Zihuatanejo, Gro., Mexico, October 27-31, 2025.  \\~\\Volume Editors: }

\SetBgColor{textcolor}
\SetBgOpacity{1}
\SetBgAngle{90}
\SetBgPosition{current page.center}
\SetBgVshift{-0.46\textwidth}
\SetBgScale{1.2}
\SetBgContents{\sffamily\Text}

\corres{ejarvela@ttu.edu}

\received{\today}
\accepted{\today}

\setbool{rho-abstract}{true} 
\setbool{rho-resumen}{true} 

\begin{resumen}
\textbf{To be translated:} We report the discovery of extreme 37~GHz radio variability in early-stage active galactic nuclei (AGN). Most of these sources belong to the class of narrow-line Seyfert 1 (NLS1) galaxies, which harbour fast-growing, low-mass supermassive black holes that accrete at high Eddington ratios. At 37~GHz, these extraordinary AGN exhibit amplitude variability of three to four orders of magnitude over timescales of a few days. Notably, despite several attempts, we have not detected relativistic jets in these sources, suggesting that this phenomenon may occur very close to the black hole and may represent a new form of AGN variability. One of these sources, which flared at 37~GHz, was followed up with the Karl G. Jansky Very Large Array and \textit{Swift} XRT/UVOT. Based on these observations, we estimated an $e$-folding timescale of mere hours, leading to variability brightness temperatures and variability Doppler factors that are extremely difficult to explain by an incoherent emitter. This suggests a possible detection of coherent emission from an AGN, although more observations will be needed to distinguish between alternative scenarios.
\end{resumen}

\begin{abstract}
We report the discovery of extreme 37~GHz radio variability in early-stage active galactic nuclei (AGN). Most of these sources belong to the class of narrow-line Seyfert 1 (NLS1) galaxies, which harbour fast-growing, low-mass supermassive black holes that accrete at high Eddington ratios. At 37~GHz, these extraordinary AGN exhibit amplitude variability of three to four orders of magnitude over timescales of a few days. Notably, despite several attempts, we have not detected relativistic jets in these sources, suggesting that this phenomenon may occur very close to the black hole and may represent a new form of AGN variability. One of these sources, which flared at 37~GHz, was followed up with the Karl G. Jansky Very Large Array and \textit{Swift} XRT/UVOT. Based on these observations, we estimated an $e$-folding timescale of mere hours, leading to variability brightness temperatures and variability Doppler factors that are extremely difficult to explain by an incoherent emitter. This suggests a possible detection of coherent emission from an AGN, although more observations will be needed to distinguish between alternative scenarios.
\end{abstract}

\begin{document}

\maketitle
\pagestyle{fancy}
\thispagestyle{firststyle}


\section{Introduction}

\RMxAAstart{A}ctive galactic nuclei (AGN), powered by accretion onto a supermassive black hole, are among the most luminous non-transient sources in the Universe. Their basic structure is rather similar, but only $\sim$10\% of them can host powerful relativistic jets \citep{2016padovani1} and exhibit substantial variability, especially pronounced in radio and gamma-rays, caused by changes in the jet. Only a few AGN classes host these jets. These include blazars --- flat-spectrum radio quasars (FSRQs) and BL Lacs --- whose jets we observe at small angles; their misaligned radio galaxy counterparts; and a class of early-stage, fast-accreting AGN, called narrow-line Seyfert 1 (NLS1) galaxies. FSRQs and BL Lacs are well-studied, fully evolved AGN that we know differ in accretion modes and environments \citep{2012best1}. On the other hand, whereas earlier observations hint at the presence of relativistic jets in NLS1s \citep{1991remillard1}, their existence was confirmed $\sim$20 years ago \citep{2006komossa1,2009abdo2}, and several aspects of them remain poorly understood.

NLS1s are identified based on their optical spectra; their full-width at half maximum (FWHM) of H$\beta$ $<$ 2000~km s$^{-1}$ \citep{1985osterbrock1}, and the flux ratio [O~III] / H$\beta$ $<$ 3. They also often show strong Fe~II emission \citep{1989goodrich1}. The narrow H$\beta$ implies that the black hole masses in NLS1s are low ($<10^8 M_\odot$, \citealp{2018komossa1}), as confirmed by reverberation mapping \citep{2016wang1}. However, their luminosities are comparable to those of broad-line Seyfert 1 galaxies, thereby yielding very high Eddington ratios, often exceeding unity. Their black hole masses and predominantly disk-like hosts \citep[e.g.,][]{2018jarvela1, 2022varglund1} suggest that they are early-stage AGN \citep{2000mathur1}. Their large-scale environments support this hypothesis; NLS1s reside in regions that, in the framework of cosmic downsizing, develop later than dense parts, such as superclusters, where most blazars reside \citep{2017jarvela1}. NLS1s are sources that were not expected to host relativistic jets \citep{2000laor1}; thus, the discovery of powerful relativistic jets in NLS1s \citep{2006komossa1, 2009abdo2} broke the conventional jet paradigm in which only the most massive black holes are able to launch relativistic jets \citep{2000laor1}.

A few per cent of NLS1s host relativistic jets \citep{2006komossa1,2016lister1}; in contrast, most NLS1s (85\%) have never been detected in radio, or are weak radio sources (10\%) in surveys such as the Faint Images of the Radio Sky at Twenty-Centimeters. According to a recent study of a large sample of NLS1s \citep{2020berton1}, the detection fraction may be even lower \citep[7-8\%,][]{2025varglund1}. NLS1s exhibit diverse radio morphologies, as confirmed by arcsec-scale surveys using the Karl G. Jansky Very Large Array (JVLA) and the Australia Telescope Compact Array \citep[ATCA;][]{2018berton1,2020chen1,2022chen1}, and the origin of their radio emission remains a complex issue. Examining 44 NLS1s using 5~GHz arcsec-scale spatially resolved radio spectral index maps, \citet{2022jarvela1} reported that these NLS1s range from jet-dominated to totally host-dominated, and everything in between.

In this paper, we summarise the still-unfolding case of unexpected high-frequency radio variability observed in a small sample of NLS1s. We introduce the sample, the original discovery, and the first follow-up observations in Sect.~\ref{sec:searchofjets}. Sect.~\ref{sec:j1228} presents preliminary results for a multiwavelength follow-up of one of the sources, and Sect.~\ref{sec:conclude} discusses some future directions. Throughout this paper, we adopt a standard $\Lambda$CDM cosmology, with a Hubble constant $H_0$ = 72 km~s$^{-1}$ Mpc$^{-1}$ and $\Omega_{\Lambda}$ = 0.73.

These studies are part of the project NOCTURNE\footnote{www.nls1.space}, which stands for \textbf{N}arrow-line Seyfert 1 galaxies \textbf{O}ver \textbf{C}osmic \textbf{T}ime: \textbf{U}nification, \textbf{R}eclassification, \textbf{N}ature, and \textbf{E}volution. NOCTURNE aims to improve our overall understanding of the population of undermassive black holes, such as NLS1s, by embracing a panchromatic approach to investigating early-stage AGN.

\section{Unexpected discoveries} \label{sec:searchofjets}

\begin{figure*}[ht]
    \centering
    \includegraphics[width=1.99\columnwidth]{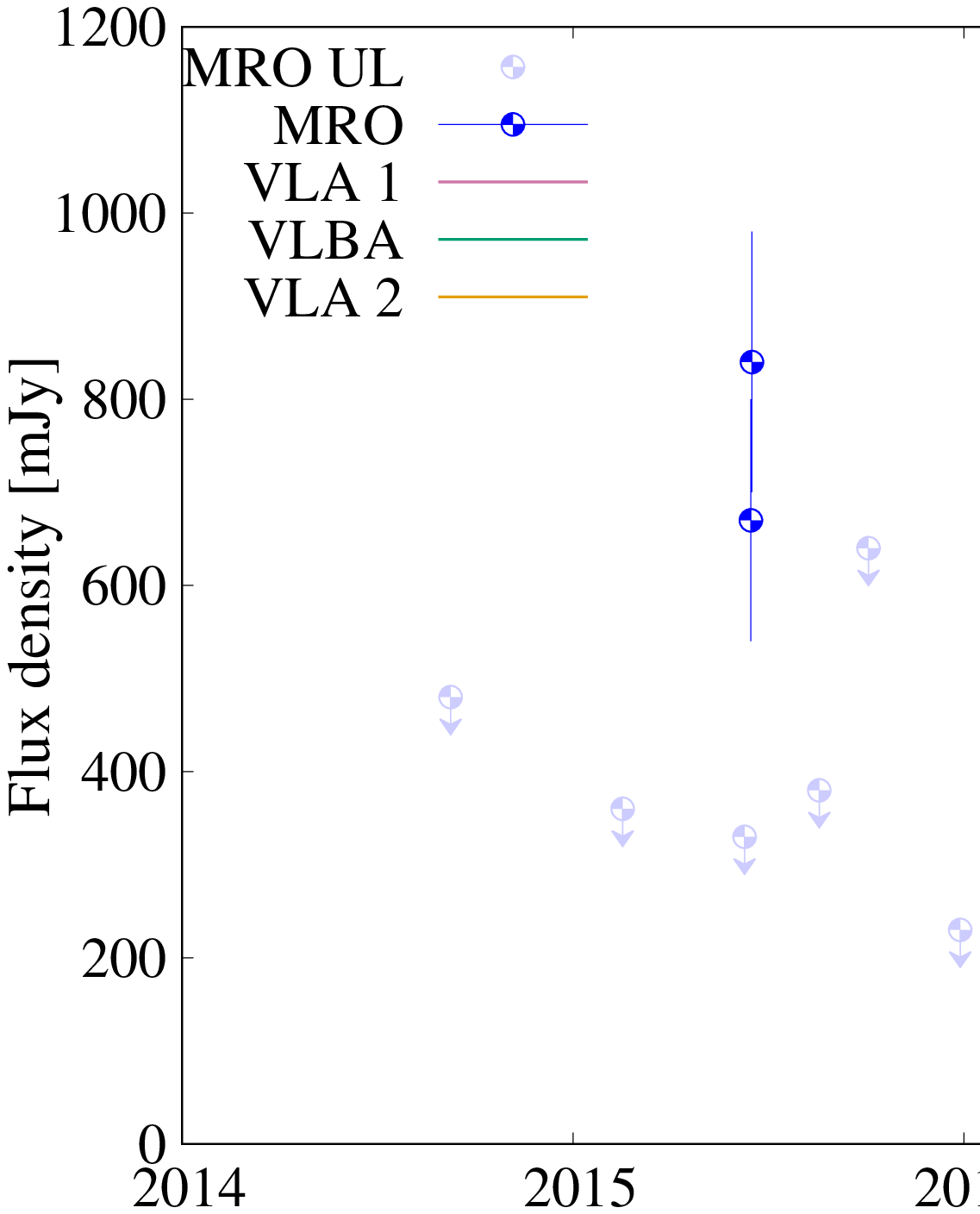}
    \caption{37~GHz MRO light curve of J1509+6137. Symbols and colours are defined in the legend. Symbols with a downward arrow indicate upper limits. Figure from \citet{2024jarvela1}.}
    \label{fig:J1509lc}
\end{figure*}

To exploit the strong manifestation of the jets in the radio regime, we started a 37~GHz NLS1 monitoring programme at the Metsähovi Radio Observatory (MRO, Finland) in 2012 \citep{2017lahteenmaki1}. The aim was to identify more jetted NLS1s to start characterising the jetted fraction of the NLS1 population. Two NLS1 samples were chosen for monitoring based on properties other than radio: one was selected based on very dense Mpc-scale environments \citep{2017jarvela1}, and the other based on spectral energy distributions that appeared favourable for 37~GHz observations. Surprisingly, we discovered seven radio-weak NLS1s flaring at Jy-levels at 37~GHz \citep{2018lahteenmaki1}. The basic properties of the sample are shown in Table~\ref{tab:sample}, and the 37~GHz light curve of one of these sources, J1509+6137, is shown in Fig.~\ref{fig:J1509lc}. From the light curve, it is clear that these sources exhibit flares similar to those observed in blazars, strongly implying that they host relativistic jets. Moreover, one source, J1641+3454, was identified as a gamma-ray emitter, and two others had test statistic (TS) values $>$ 20, supporting the interpretation that relativistic jets are responsible for the flares. 

\begin{table}[ht]
\caption{Basic properties of the sample.}
 \label{tab:sample}
 \centering
 \begin{tabular*}{.99\columnwidth}{@{}l@{\hspace*{12pt}}l@{\hspace*{12pt}}l@{\hspace*{16pt}}l@{}}
  \hline
  Name & RA & Dec & $z$ \\
    & (hh mm ss.ss) & (dd mm ss.ss) & \\  \hline
J1029+5556 & 10 29 06.69 & +55 56 25.25 & 0.451 \\ 
J1228+5017 & 12 28 44.82 & +50 17 51.24 & 0.262 \\
J1232+4957 & 12 32 20.12 & +49 57 21.82 & 0.262 \\
J1509+6137 & 15 09 16.17 & +61 37 16.80 & 0.201 \\
J1510+5547 & 15 10 20.05 & +55 47 22.11 & 0.150 \\
J1522+3934 & 15 22 05.50 & +39 34 40.45 & 0.077 \\
J1641+3454 & 16 41 00.10 & +34 54 52.67 & 0.164 \\
  \hline
 \end{tabular*}
\end{table}

\begin{figure}[ht]
    \centering
    \includegraphics[width=0.99\columnwidth]{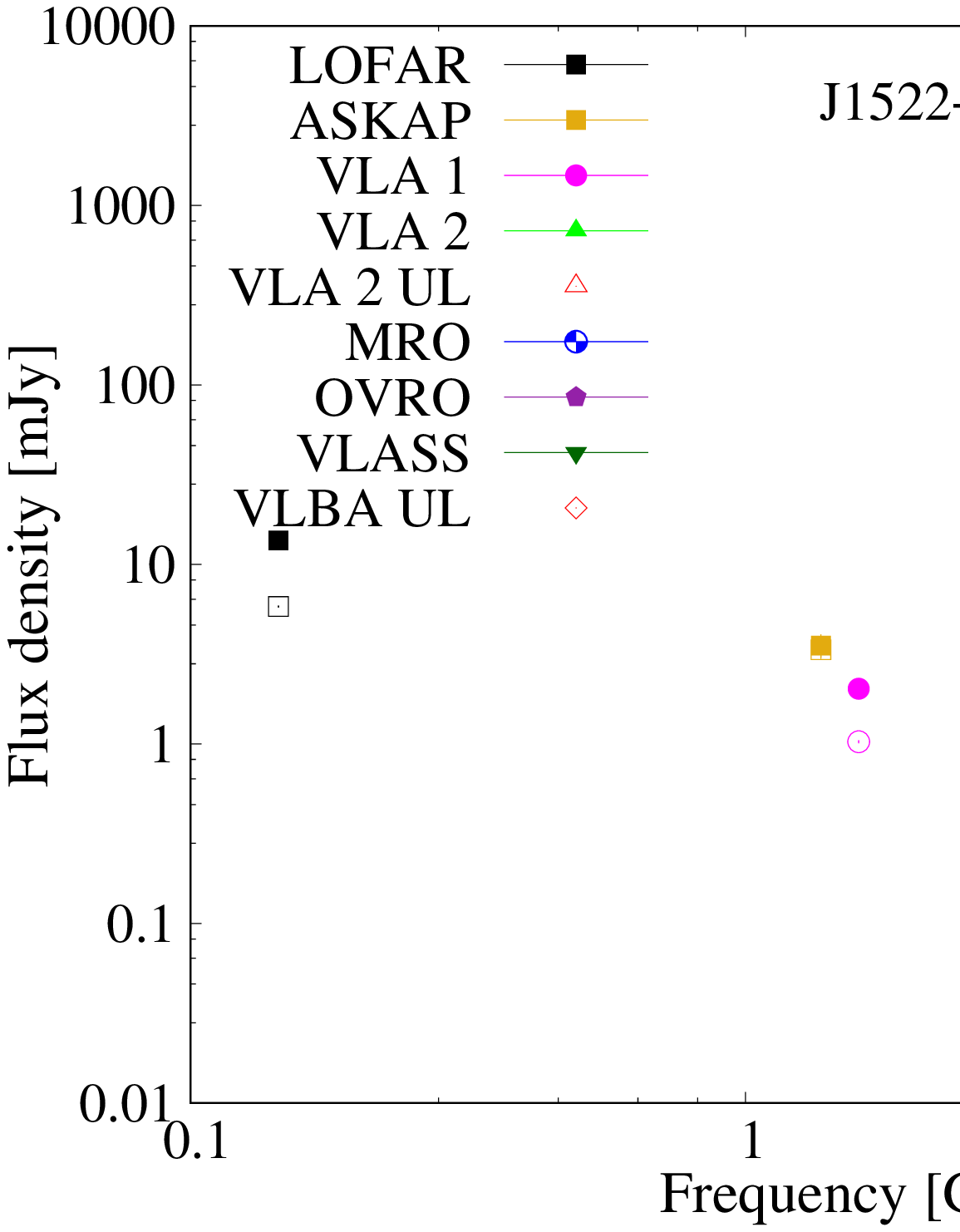}
    \caption{Non-simultaneous radio spectrum of J1522+3934. Symbols and colours are defined in the legend. Filled symbols mark integrated flux densities, empty symbols peak flux densities, and empty symbols with a downward arrow upper limits. Figure from \citet{2024jarvela1}.}
    \label{fig:J1522spectrum}
\end{figure}

To confirm the presence of jets, we conducted follow-up radio imaging observations with the JVLA from 1.6 to 45~GHz, and with the Very Long Baseline Array (VLBA) at 15~GHz \citep{2020berton2, 2024jarvela1}. Additionally, the Owens Valley Radio Observatory (OVRO, USA) began monitoring these sources at 15~GHz in 2020. However, despite these attempts, no jets were detected. The VLBA observations yielded only non-detections (rms $\sim$60-70~$\mu$Jy), and in the JVLA observations, all sources were weak, as can be seen in Fig.~\ref{fig:J1522spectrum}, which shows the non-simultaneous radio spectrum of one of these sources, J1522+3934. Two distinct states characterise the spectrum; in the low state, the spectrum is steep and consistent with optically thin synchrotron emission or, alternatively, with star formation activity. Whereas in the high state, the spectrum turns strongly inverted, indicating extreme variability. We emphasise that the OVRO observations at 15~GHz are not simultaneous with the MRO observations at 37~GHz; however, the spectrum indicates that the flux densities increase toward higher frequencies.

To further characterise the flares, we estimated their $e$-folding timescales \citep{1999valtaoja1} using two consecutive MRO detections within a week, and thus likely from the same flare. The resulting timescales were on the order of days, whereas typical blazar timescales are weeks to months \citep{2009hovatta1}. The $e$-folding timescale also allowed us to estimate the variability brightness temperatures and variability Doppler factors of the flares, both of which are comparable to those in blazars. 

The high amplitudes and short timescales, coupled with the absence of detectable jets, already rule out several explanations for their extraordinary behaviour, including standard relativistic jets, newborn jets, geometrical effects, tidal disruption events, and extreme scattering events. After careful scrutiny of numerous possible explanations \citep[see][]{2024jarvela1}, a few remain plausible.

In one scenario, the flares would be produced by interaction between a weak jet or a jet base and a broad line region (BLR) cloud or a star. The duty cycle of such events should be very high \citep[10-100, see][]{2019delpalacio1}, and the resulting flares should last from less than a day to a few days. These flares should be observable over the whole electromagnetic spectrum.

Another scenario requires a very high BLR covering factor and plenty of ionised gas; in this case, the jet is present, but its emission is totally free-free absorbed most of the time. The flares would be produced by gaps in the BLR clouds through which the jet emission could occasionally escape. The timescale of these flares can be arbitrarily short, as it depends only on the size of the gap. Some support for this hypothesis was found in X-ray observations of J1641+3454: no absorption was detected in X-rays just after a flare, when the nucleus would have been exposed, but a possible warm absorber is seen in the X-ray spectrum when the source is in a low state \citep{2023romano1}.

The third explanation evokes magnetic reconnection in the black hole magnetosphere as a possible source of the flares \citep[e.g.,][]{2022ripperda1,2022kimura1}. Its exact observational signatures in the radio regime are not well-defined yet, but based on the model by \citet{2015kadowaki1}, an effectively accreting black hole with a mass of 10$^{7} M_{\sun}$ and turbulence-induced fast reconnection can produce magnetic reconnection power spanning from 10$^{39}$ to 10$^{43}$ erg s$^{-1}$ --- enough to explain the flares in our sources.

It should also be noted that the non-detection of jets does not conclusively rule out their presence. Weak or intermittent jets with a favourable geometry and possibly precession could perhaps produce these flares. 

\section{Multiwavelength view} \label{sec:j1228}

\begin{figure*}[ht]
    \centering
    \includegraphics[width=1.88\columnwidth]{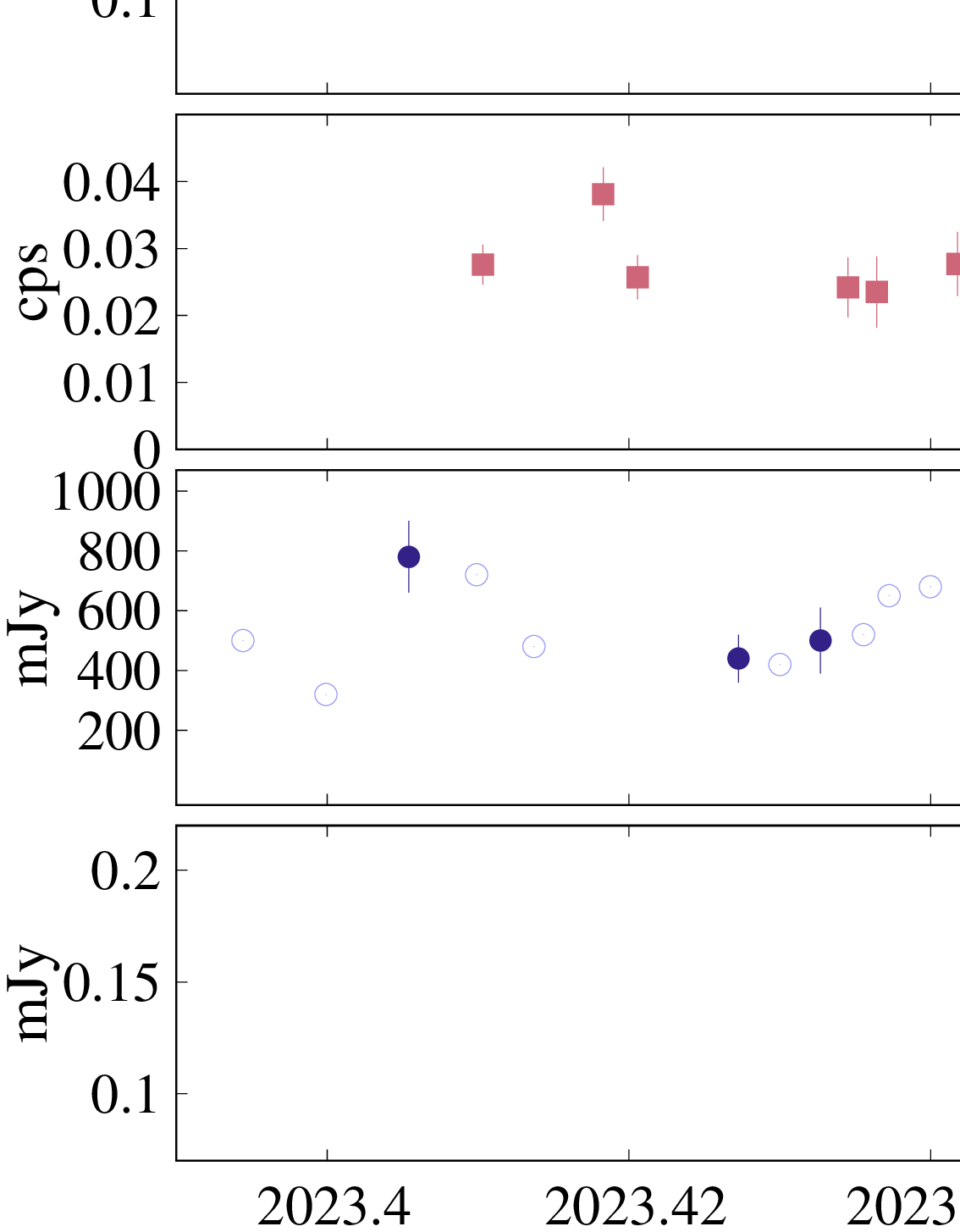}
    \caption{Multiwavelength light curve of J1228+5017. The panels from top to bottom are: \textit{Swift} UVOT V-band, \textit{Swift} XRT 0.3-10~keV, MRO 37~GHz, and JVLA Ka-band. Filled symbols denote detections, and empty symbols indicate upper limits.}
    \label{fig:J1228lc}
\end{figure*}

These scenarios should exhibit distinct multiwavelength signatures, particularly at high energies, such as X-rays, that can be used to distinguish among them. Possibly correlated radio and X-ray variability has earlier been observed in one source \citep{2023romano1}. Thus, when another source, J1228+5017, flared several times at 37~GHz, we launched a multiwavelength effort to follow it up with the JVLA and \textit{Swift} XRT and UVOT. Light curves in selected bands, \textit{Swift} UVOT V-band and XRT 0.3-10~keV, MRO 37~GHz, and JVLA Ka-band, are shown in Fig.~\ref{fig:J1228lc}. 

In this case, we cannot identify any clear trends between different bands. All light curves show variability by a factor of $\sim$2-3. However, due to numerous non-detections in the MRO light curve, we cannot determine the actual level of variability at 37~GHz. We can estimate the lower limit for the variability based on OVRO non-detections and resulting upper limits at 15~GHz (private communication), assuming a flat spectral index between 15 and 37~GHz. The lowest OVRO upper limit during this period was 7.4~mJy. Compared with the highest MRO flux density, this yields a variability factor of 118$\pm$23. If no flux is resolved out in Ka-band, we can compare MRO and JVLA flux densities for close detections, which would suggest variability by several orders of magnitude in all cases.

Following Eqs.~4-6 in \citet{2024jarvela1}, we can estimate the $e$-folding timescale ($\tau$), variability brightness temperature ($T_{\mathrm{b, var}}$), and variability Doppler factor ($\delta_{\mathrm{var}}$) using close detections. In one case, we have JVLA and MRO detections separated by only nine hours. These two detections give $\tau$ = 1.1088 $\substack{+0.0615 \\ -0.0509}$~h, $T_{\mathrm{b, var}}$ = 4.20 $\substack{+1.53 \\ -1.35}$ $\times$ 10$^{18}$~K, and $\delta_{\mathrm{var}}$ = 438 $\substack{+48 \\ -53}$. These values are much more extreme than what is seen in blazars, supporting the hypothesis that we are seeing a new form of AGN variability. These observations do not allow us to determine the origin of these flares conclusively, but based on the timescale, we can possibly rule out the scenario in which the jet or jet base interacts with a star or a BLR cloud. 



\section{Summary} \label{sec:conclude}

It appears that the multiwavelength behaviour of the extreme radio flares is not straightforward to understand. Given the very short timescales in question, it might be that with sparse sampling, we are missing the flares in other bands. Alternatively, these flares may manifest strongly only in the radio band. One option for such a phenomenon is coherent emission. It can naturally arise in the magnetised plasma around a supermassive black hole, for example, in relativistic magnetised shocks induced by the disk wind, magnetic reconnection, or changes in the jets \citep{2021sironi1}. Coordinated triggered, or preferably monitoring observations, across the electromagnetic spectrum will be required to pin down the multiwavelength characteristics of these flares.

In radio, what will be needed to decipher this extraordinary behaviour is better temporal and frequency resolution, as well as full-polarisation observations. Higher cadence will allow us to observe full flares rather than just peaks and improve the timescale estimates. With better frequency resolution, we can establish whether the flares are narrow- or broad-band, potentially distinguishing between coherent and incoherent emission. The polarisation information can further help us narrow down the alternatives, as we, for example, expect coherent emission to be extremely polarised. One future facility capable of these observations would be the next-generation Very Large Array.


\bibliography{rmxac}


\end{document}